# Design and Simulation of a Steam Turbine Generator using Observer Based and LQR Controllers

Mustefa Jibril [1], Messay Tadese [2], Eliyas Alemayehu Tadese [3]

[1.] School of Electrical & Computer Engineering, Dire Dawa Institute of Technology, Dire Dawa, Ethiopia
[2.] School of Electrical & Computer Engineering, Dire Dawa Institute of Technology, Dire Dawa, Ethiopia
[3.] Faculty of Electrical & Computer Engineering, Jimma Institute of Technology, Jimma, Ethiopia
mustefazinet1981@gmail.com

**Abstract:** Steam turbine generator is an electromechanical system which converts heat energy to electrical energy. In this paper, the modeling, design and analysis of a simple steam turbine generator have done using Matlab/Simulink Toolbox. The open loop system have been analyzed to have an efficiency of 76.92 %. Observer based & linear quadratic regulator (LQR) controllers have been designed to improve the generating voltage. Comparison of this two proposed controllers have been done for increasing the performance improvement to generate a 220 Dc volt. The simulation result shows that the steam turbine generator with observer based controller has a small percentage overshoot with minimum settling time than the steam turbine generator with LQR controller and the open loop system. Finally, the steam turbine generator with observer based controller shows better improvement in performance than the steam turbine generator with LQR controller.

## 1. Introduction

A Steam Turbine is a mechanical system that takes thermal energy from pressurized steam and converted it into mechanical output. Because the turbine generates rotational speed motion, it is particularly used to drive electrical generators. a steam turbine gain power from a steam. As hot, steam passes through the turbine ' the blades will spinning. Steam expands and cools, convert most of the energy it contains. This steam continuously spins the blades. The blades converts the potential energy of the steam into kinetic energy. Steam turbines is used to drive a generator in order to produce electricity. In this paper, a steam turbine is designed to run a series wound DC generator. The shaft of the turbine is connected to the shaft of the generator with two gears at each side with a gear ratio between them.

## 2. Mathematical Modeling

Figure 1 shows the diagram of the steam turbine with a DC generator interconnection unit.

The mechanical power developed by the steam turbine expressed by the equation:

$$\frac{dW}{dt} = V\frac{d\rho}{dt} = F_{in}(t) - F_{out}(t) \qquad (1)$$

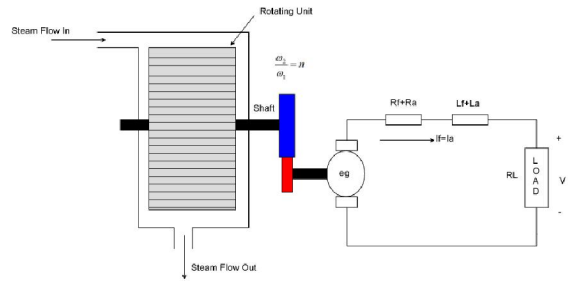

Figure 1. Steam turbine generator

Where:
$W$ weight of steam in the steam turbine [g];
$V$ – Volume of steam turbine [m3];
$\rho$ – Density of steam [g/cm3];
$F$ – Steam mass flow rate [g/s];
$t$ – Time [sec.].

Assuming that the output mass flow rate of the steam turbine is proportional to the pressure in the turbine system:

$$F_{out} = P\frac{F_0}{P_0} \qquad (2)$$

Where:
$P$ – Steam pressure in the system [kPa];
$P0$ – rated pressure;

$F0$ – rated output mass flow of the system

Assuming constant temperature in the system:

$$\frac{d\rho}{dt} = \frac{dP}{dt} \bullet \frac{\partial \rho}{\partial P} \quad (3)$$

From equations (1) to (3), it results:

$$F_{in}(t) - F_{out}(t) = V \frac{dP}{dt} \bullet \frac{\partial \rho}{\partial P} = V \bullet \frac{\partial \rho}{\partial P} \bullet \frac{dF_{out}}{dt} = \tau_T \frac{dF_{out}}{dt}$$

Where

$$\tau_T = V \bullet \frac{\partial \rho}{\partial P}$$ - Time constant

Then Equation (1) become

$$\tau_T \frac{dF_{out}(t)}{dt} + F_{out}(t) = F_{in}(t) \quad (4)$$

The steam mass flow rate is

$$F(t) = F_{in}(t) - F_{out}(t)$$

The turbine speed is integrally related to the steam flow rate:

$$\omega_1(t) = \int F(t) = \int (F_{in}(t) - F_{out}(t))dt \quad (5)$$

Substituting Equation (5) for $F_{out}(t)$ into Equation (4) yields the transfer function between the input steam mass flow and the output turbine torque.

$$\frac{\omega_1(s)}{F_{in}(s)} = \frac{\tau_T}{\tau_T s + 1} \quad (6)$$

**2.1 Modeling of DC Generator**

The generated voltage is directly proportional to the generator input speed

$$e_g = k_1 \omega_2 \quad (7)$$

Where: $k1$ is a speed proportional constant.

The generator input speed is the product of the gear ratio and the steam turbine output speed

$$\omega_2 = n\omega_1 \quad (8)$$

Therefore Equation (7) becomes

$$e_g = k_1 n \omega_1 \quad (9)$$

Then the transfer function between $F_{in}(s)$ and $e_g(s)$ become

$$\frac{e_g(s)}{F_{in}(s)} = \frac{k_1 n \tau_T}{\tau_T s + 1}$$

The equation of the generator is

$$e_g(t) = (L_f + L_a)\frac{di_a(t)}{dt} + (R_f + R_a + R_L)i_a(t) \quad (10)$$

Now substituting Equation (9) in to Equation (10) yields

$$k_1 n \omega_1(t) = (L_f + L_a)\frac{di_a(t)}{dt} + (R_f + R_a + R_L)i_a(t) \quad (11)$$

Taking the Laplace transform and obtaining the transfer function the input speed and the output current yields

$$\frac{I_a(s)}{\omega_1(s)} = \frac{nk_1}{s(L_f + L_a) + (R_f + R_a + R_L)} \quad (12)$$

The output voltage is simply the product of load resistance times the armature current

$$V(s) = R_L I_a(s) \quad (13)$$

Now substituting Equation (13) in to Equation (12) yields

$$\frac{V(s)}{\omega_1(s)} = \frac{nR_L k_1}{s(L_f + L_a) + (R_f + R_a + R_L)} \quad (14)$$

Finally combining Equation (6) to Equation (14) yields to the transfer function between the input steam mass flows to the output generated voltage

$$\frac{V(s)}{F_{in}(s)} = \frac{nR_L k_1 \tau_T}{(\tau_T s + 1)(s(L_f + L_a) + (R_f + R_a + R_L))}$$

The parameters of the system is shown in Table 1 below

Table 1 System parameters

| No | Parameters | Symbol | Value |
|---|---|---|---|
| 1 | Time constant | $\tau_T$ | 2 second |
| 2 | Speed proportional constant. | $k_1$ | 4 Vs/rad |
| 3 | Gear ratio | $n$ | 4 |
| 4 | Field winding inductance | $L_f$ | 3 H |
| 5 | Field winding resistance | $R_f$ | 2 ohm |
| 6 | Armature winding inductance | $L_a$ | 4 H |
| 7 | Armature winding resistance | $R_a$ | 4 ohm |
| 8 | Load resistance | $R_L$ | 8 ohm |

Numerically the transfer function become

$$\frac{V(s)}{F_{in}(s)} = \frac{18}{s^2 + 2.5s + 1}$$

The state space form of the transfer function is

$$\dot{x} = \begin{pmatrix} -2.5 & -1 \\ 1 & 0 \end{pmatrix} x + \begin{pmatrix} 1 \\ 0 \end{pmatrix} u$$

$$y = \begin{pmatrix} 0 & 18 \end{pmatrix} x$$

## 3. The Proposed Controllers Design
### 3.1 Observer-Based Controller Design

The deal with the general case where only a subset of the states, or linear combinations of them, are obtained from measurements and are available to our controller. Such a guidelines is referred to as the output feedback problem.

The output is of the form

$$y = Cx + Du \quad (15)$$

We shall examine a class of output feedback controllers constructed in two stages:

1. Contracting an observer | a system dynamics that is driven by the inputs and the outputs of the system, and yield an deliberation of its state variables;
2. Using the estimated state instead of the actual state in a estate response scheme.

The block diagram of the steam turbine generator system with the observer-based controller is shown in Figure 2 bellow.

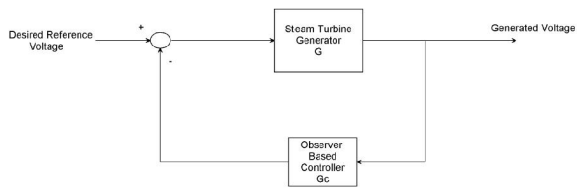

Figure 2 Block diagram of the steam turbine generator system with the observer-based controller

The controller $G_c(s)$ can be further derived in the following form:

$$G_c(s) = I - K(sI - A + BK + HC)^{-1} B \quad (16)$$

With its state space realization

$$G_c(s) = \begin{bmatrix} A - BK - HC & B \\ -K & I \end{bmatrix} \quad (17)$$

The controller $G_c(s)$ in Equation (17) is called the observer-based controller, since the structural idiot of the observer is implicitly reflected within the controller.

Where the state space model of the plant, G, the state feedback gain vector K, and the observer gain vector H are then returned, respectively.

We select the weighting matrix Q and R as

$$Q = \begin{pmatrix} 8 & 0 \\ 0 & 8 \end{pmatrix} \quad and \quad R = 1$$

And we select the observer gain vector as

$$H = \begin{pmatrix} 2 \\ -0.5 \end{pmatrix}$$

And we obtain the state feedback gain vector K as

$$K = \begin{bmatrix} 1.7720 & 2 \end{bmatrix}$$

The observer-based controller state space representation become

$$\dot{x}_o = \begin{pmatrix} -4.272 & -39 \\ 1 & 9 \end{pmatrix} x + \begin{pmatrix} 2 \\ -0.5 \end{pmatrix} u_o$$

$$y_o = (1.772 \quad 2) x_o$$

### 3.2 LQR Controller Design

LQR is a controller that gives the best possible achievement with respect to some given measure of performance. The realization measure is a quadratic function composed of state vector and manipulated input. The block diagram of the steam turbine generator with LQR controller is shown in Figure 2 below.

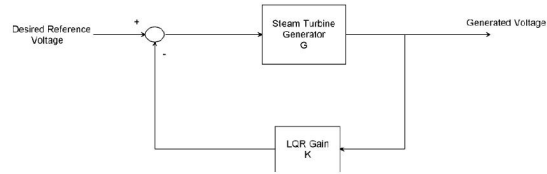

Figure 3 Block diagram of the steam turbine generator with LQR controller

Here we choose Q and R matrixes as

$$Q = \begin{pmatrix} 3 & 0 \\ 0 & 3 \end{pmatrix} \quad and \quad R = 5$$

The value of obtained feedback gain matrix K of LQR is given by

$$K = \begin{bmatrix} 0.2166 & 0.2649 \end{bmatrix}$$

## 4. Result and Discussion

In this section, the open loop simulation of the generated voltage and power, the simulation of the performance comparison of the proposed controllers for generating a desired voltage is discussed below.

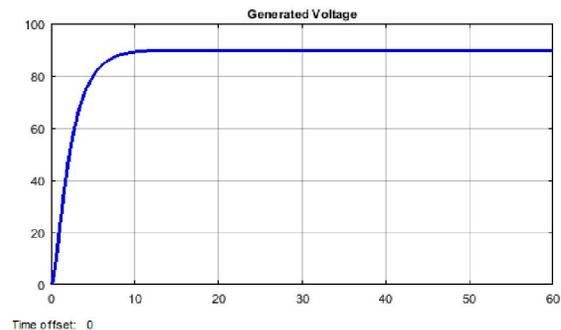

Figure 4. Open loop generated voltage

### 4.1 Simulation of the open loop steam turbine generator voltage and power

The open loop voltage and power generated for a steam mass flow of 5 g/s input is shown in Figure 4 and Figure 5 respectively.

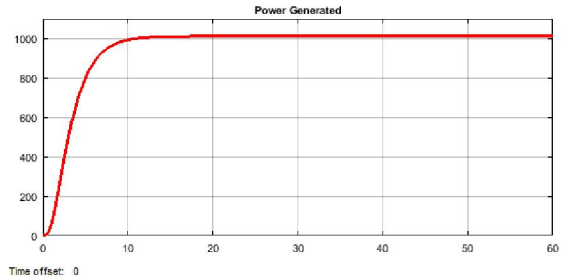

Figure 5. Open loop output power

As Figure 4 and Figure 5, in the first 10 seconds, the voltage increases from 0 to 90 volts and it reaches is steady state value of 90 volt and the output power is 1000 W. The induced voltage in the armature circuit and the input power to the generator simulation result is shown in Figure 6 and Figure 7 respectively.

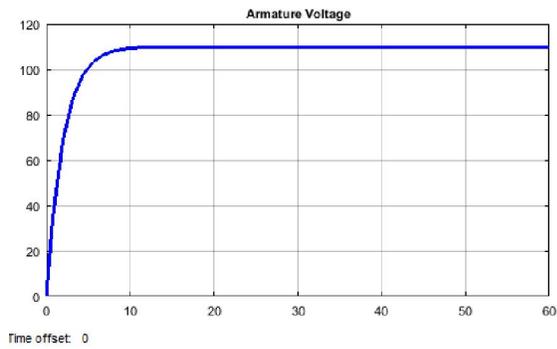

Figure 6. The induced voltage in the armature circuit

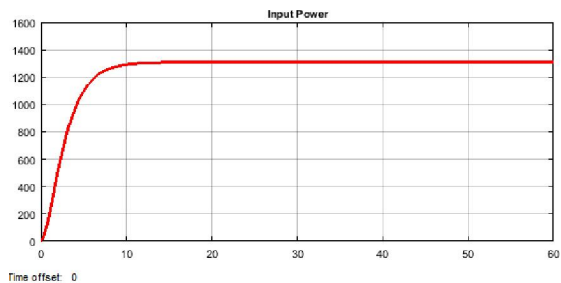

Figure 7. Input power

The efficiency of the system become

$$\frac{Output\ power}{Input\ power} \times 100\% = \frac{1000\,W}{1300\,W} \times 100\% = 76.92\%$$

### 4.2 Comparison of the Steam Turbine Generator with Observer Based & LQR Controllers for Generating a Desired Voltage

The simulation result of the performance of the steam turbine generator with observer based & LQR controllers for generating a desired voltage of 220 Dc volt is shown in Figure 8 below.

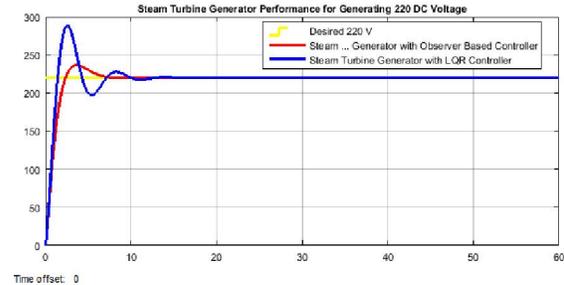

Figure 8. Performance of the steam turbine generator with observer based & LQR controllers for generating a desired voltage of 220 Dc volt

As seen from Figure 8, the steam turbine generator with observer based controller has a small percentage overshoot and a settling time of 7 second which is better than the steam turbine generator with LQR controller and the open loop system.

### 5. Conclusion

In this paper, a simple steam turbine generator is modeled and designed using a series wound Dc generator. The open loop system is simulated for generated voltage, input and output power. For a steam mass flow of 5 g/s the steam turbine generator generates a 90 volt with an output power of 1000 W. The generator induced armature circuit voltage is simulated with an input power of 1300 W. The efficiency of the simple steam turbine generator is 76.92 %. Comparison of the steam turbine generator with observer based & LQR controllers for generating a desired voltage of 220 Dc volt have been simulated and the simulation result shows that the steam turbine generator with observer based controller has a small percentage overshoot and less settling time than the steam turbine generator with LQR controller and the open loop system. Finally the comparison and analysis result prove the effectiveness of the proposed steam turbine generator with observer based controller.

6/20/2020